# VTS-Guided AI Interaction Workflow for Business Insights

Sun Ding
ding@scantist.com

Ude Enebeli
Ude.enebeli@mtn.com

Atilhan(Ati) Manay
ati.manay@rheem.com

Ryan Pua
ryan_pua@singaporeair.com.sg

Kamal Kotak
kamalkotak1976@gmail.com

## Abstract

Modern firms face a flood of dense, unstructured reports. Turning these documents into usable insights takes heavy effort and is far from agile when quick answers are needed. VTS-AI tackles this gap. It integrates Visual Thinking Strategies, which emphasize evidence-based observation, linking, and thinking, into AI agents, so the agents can extract business insights from unstructured text, tables, and images at scale. The system works in three tiers (micro, meso, macro). It tags issues, links them to source pages, and rolls them into clear action levers stored in a searchable YAML file. In tests on an 18-page business report, VTS-AI matched the speed of a one-shot ChatGPT prompt yet produced richer findings: page locations, verbatim excerpts, severity scores, and causal links. Analysts can accept or adjust these outputs in the same IDE, keeping human judgment in the loop. Early results show VTS-AI spots the direction of key metrics and flags where deeper number-crunching is needed. Next steps include mapping narrative tags to financial ratios, adding finance-tuned language models through a Model-Context Protocol, and building a Risk & Safety Layer to stress-test models and secure data. These upgrades aim to make VTS-AI a production-ready, audit-friendly tool for rapid business analysis.

## Introduction

In today's data-saturated enterprises, decision-makers must sift through a mosaic of financial statements, employee surveys, operational logs, and market chatter to find the handful of signals that truly shape performance. Unearthing those signals usually demands rare combinations of domain expertise, quantitative skill, and the patience to reconcile data that arrives in wildly different formats. Consequently, valuable insights often surface late, if they surface at all, after considerable cost and effort.

**Visual Thinking Strategies (VTS)** offers a lightweight alternative. Originating as an inquiry-based art pedagogy, VTS asks participants to ground every interpretation of an image in what they can actually see, using three simple prompts:

1. *"What's going on in this image?"*
2. *"What do you see that makes you say that?"*
3. *"What more can we find?"*

When those questions are transplanted into a business context, the "image" becomes a dashboard, a paragraph in an annual report, or a snippet of customer feedback. Teams learn to back each hypothesis with visible evidence, entertain multiple explanations, and converge on what matters most to the customer or the firm, mirroring best practices in data-driven product and strategy work, but without the heavy analytical overhead. Figure 1 depicts a VTS session set up in a museum gallery.

AI agents are autonomous software entities that observe an environment, reason over goals, and trigger actions, which often call external APIs or other tools, while retaining memory that lets them adapt and improve. Figure 2 gives an overview of how such agents perceive, decide, and act in single-agent, multi-agent, and human-in-the-loop configurations.

The emergence of AI agents gives organizations unprecedented capacity to scan and summarize vast troves of unstructured information, like earnings calls, policy documents, social-media threads, at a speed and consistency no human analyst can match. Marrying VTS's disciplined inquiry with the reach of these agents hints at a practical pathway from "What do we see?" to "What business value does that reveal, and how might we act?"

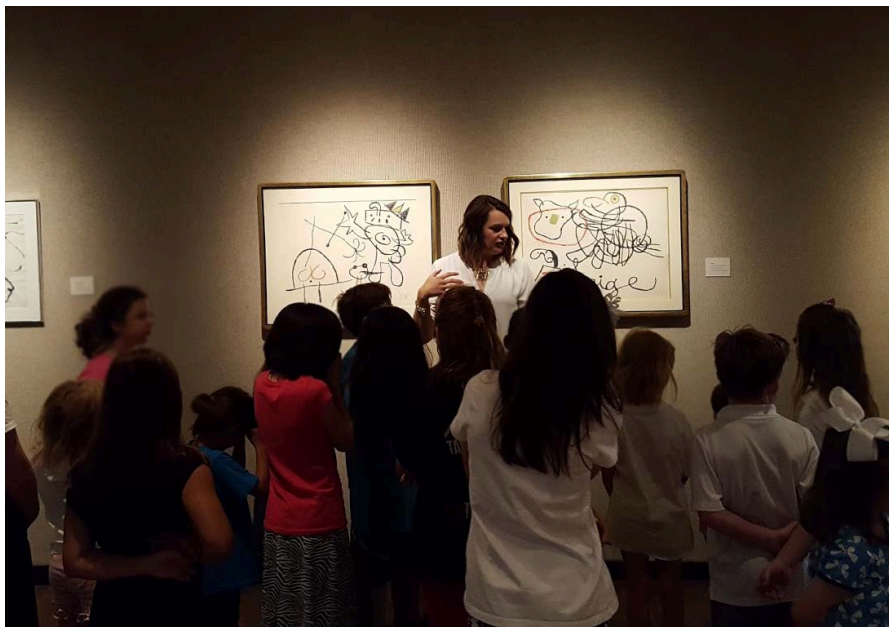

**Figure 1. Visual Thinking Strategies (VTS) uses images to give students space to observe, think, and communicate.**

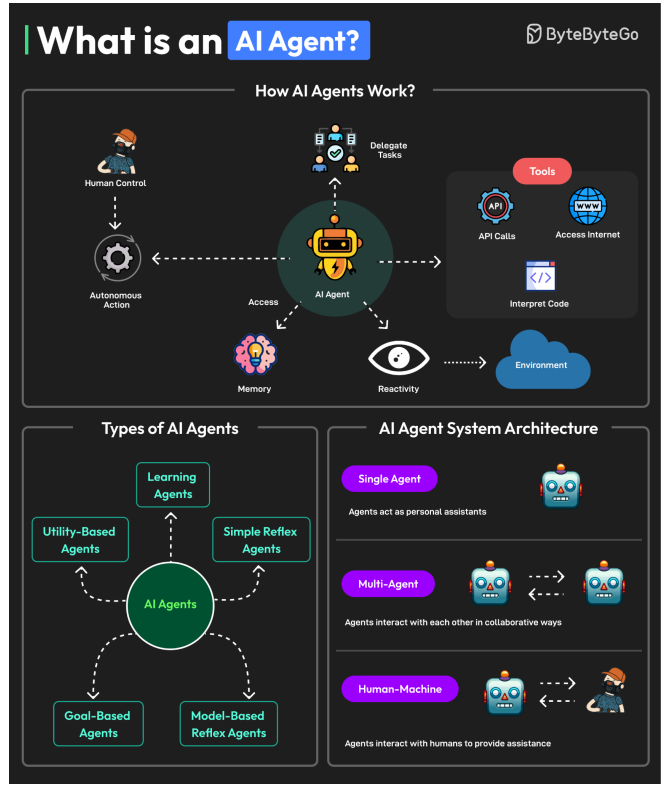

**Figure 2. AI Agents Overview, cited from ByteByteGo**

In this exploratory study we sketch how a **VTS-guided AI workflow** could adapt the three-question cycle to textual and numerical business artifacts. Our proof-of-concept illustrates how managers with different technical backgrounds might (1) capture initial impressions, (2) link those impressions to concrete evidence, and (3) iteratively refine their understanding. While our early findings are illustrative rather than conclusive, they suggest that such a "qualitative sensing" layer could democratize analytical thinking and shorten the path from raw data to informed action.

The remainder of this paper surveys related work, outlines the proposed agent design, shares observations, and maps out next steps, both for systematic evaluation and for addressing security and risk considerations in future deployments.

## Related Work

Breakthroughs in Large Language Models (LLMs) and the first wave of autonomous AI agents have prompted finance to re-imagine how it mines sprawling, unstructured reports for insights. Modern LLMs can summarize 10-K filings, label documents, answer deep questions, and even draft investment memos [1][2]. Domain-tuned variants such as BloombergGPT and JPMorgan's DocLLM push accuracy higher by mastering sector-specific jargon and table formats [3][4], while MIT studies show LLMs supporting sentiment analysis, time-series forecasting, and full-market simulations [5][6].

AI agents extend LLM power by planning tasks, calling external tools, and learning from feedback instead of replying to a single query [7]. Multi-agent systems divide labor. For example, one agent extracts KPIs, another verifies the math. The multi-agent systems raise precision on difficult numerical tasks [8]. MIT CSAIL has demonstrated agents that handle exceptions in a human-like way, a critical skill for messy real-world finance [9].

State-of-the-art solutions rest on several technical pillars:

- domain fine-tuning on massive finance corpora (e.g., Bloomberg's 345 billion tokens) for sharper jargon and figure handling [3];
- Retrieval-Augmented Generation, which grounds answers in trusted sources and leaves an audit trail [13][14];
- hybrid and multimodal setups that pair LLMs with Graph Neural Networks for fraud graphs or with OCR to read scanned tables [15];
- agentic workflows that break work into small steps, tap APIs, run Python for calculations, and preserve context [7][8];
- document-aware layouts such as DocLLM that merge page structure with text [4];
- numerical helpers that off-load exact maths to calculators or symbolic engines [11].

Large institutions already harness these advances. JPMorgan's COiN slashed contract review from hours to seconds [4]; HSBC automates AML reports with agents [12]; Ant Group boosts credit scoring with LLM analytics [2]; Wells Fargo fields LLM chatbots for customer queries [10]; and Apollo Global re-engages customers across its portfolio using agent-driven outreach [6].

These successes show what is possible, but also spotlight a divide: complex RAG pipelines, hybrid models, and multi-agent orchestration demand deep pockets, specialized talent, and proprietary data that smaller firms lack. Non-technical users, meanwhile, still distrust opaque "black-box" answers, especially when precise numbers are involved. This is because current systems rarely reveal how conclusions are formed [12]. A central question remains: **can non-technical professionals unlock the same value through low-cost yet effective prompts?** This work is driven by that challenge. Our goal is therefore to simplify interaction: to provide an intuitive prompting framework that lets everyday analysts run thoughtful, end-to-end evaluations and see exactly how each insights was reached, without needing advanced AI expertise.

## Methods

We adapt Visual Thinking Strategies (VTS) from art appreciation to business document analysis, creating a systematic three-tier framework that guides AI agents through structured inquiry. Rather than asking AI to generate immediate conclusions, we prompt agents to **observe**, **evidence**, and **explore** business documents using VTS's foundational questions.

## VTS-AI Workflow

We translate the core VTS questions for business context, as illustrated in Table 1. This creates a disciplined inquiry process where every AI insights must be grounded in observable evidence from the source documents.

**Table 1. VTS Translation**

| |
|---|
| What's going on in this document? → What business patterns/issues do you observe? |
| What do you see that makes you say that? → What specific data supports this observation? |
| What more can we find?→ What additional relationships or insights emerge? |

We apply the VTS thinking framework at three levels to capture and consolidate findings with a bottom-up approach:
- Micro Level (Individual Elements): AI agents analyze individual document pages for specific issues related to profit, performance, and employee satisfaction. Each observation is tagged with evidence location and severity scoring. This level focuses on granular detail extraction using LLM vision capabilities to process financial tables, text blocks, and visual elements.
- Meso Level (Operational Relationships): AI agents connect micro-level findings to operational parameters, identifying how specific issues relate to strategic levers (staffing, marketing, discount rates). This level performs cross-quarter analysis to track progress and establishes goal-parameter alignment mapping.
- Macro Level (Strategic Execution): AI agents analyze innovation portfolios and J-curve patterns, determining optimal project sequencing and resource allocation. This level provides execution strategy recommendations with revenue harvest timing.

We add an evidence-based validation step where AI agents must return supporting evidence after analyzing the micro, meso, and macro levels. Our implementation processes quarterly financial reports through LLM-based content extraction, converting PDFs to structured YAML files containing text, tables, and image descriptions. AI agents at each analysis level are required to provide source citations with specific page locations, direct evidence snippets from the original documents, severity ratings (High, Medium, Low) based on business impact, and explicit connections to actionable business parameters. This validation step ensures that all AI observations are traceable to source materials. This step helps our approach reduce hallucination, enable verification of insights and support transparent human-AI collaborative decision-making.

## Implementation

Our prototype implements the VTS-guided workflow as a lightweight, modular system. Ingestion, evidence extraction, reasoning, and orchestration are handled by separate components that exchange data through straightforward configuration files. This architecture allows engineers to switch to a different LLM endpoint or integrate with another enterprise data stack after only minor changes. Figure 3 depicts the pipeline at a high level, and the following sections describe each stage in sequence, from PDF rasterization to human-in-the-loop validation.

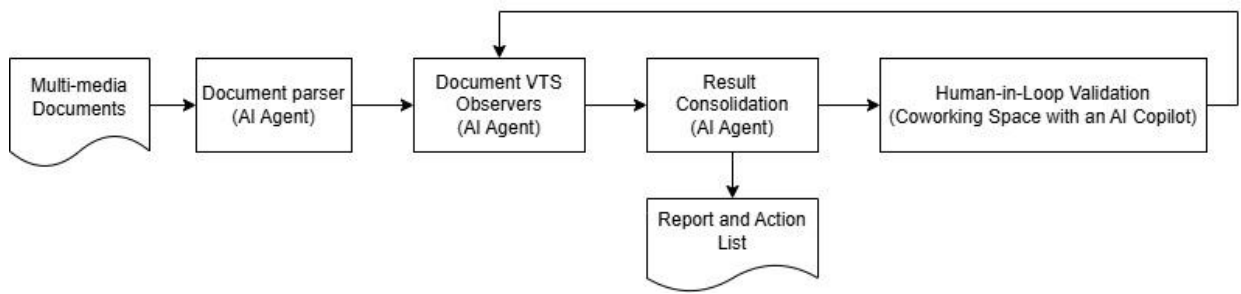

**Figure 3. VTS-AI workflow overview**

**Document Parser**

Conventional OCR frameworks proved either too brittle (open-source) or too costly (commercial SaaS) for quarterly, multi-hundred-page filings. We therefore built **pdf2image**, a two-step micro-service that (i) rasterizes each PDF page at 300 DPI in under 100 milliseconds, preserving layout for later citation, and (ii) streams the resulting PNG directly to a vision-enabled LLM (Azure GPT-4o). The model returns a YAML fragment in which every element is tagged as *text*, *table* (CSV serialized), or *figure* (caption plus bounding box).

**VTS Observers**

Three VTS observer modules, written in LangGraph (https://www.langchain.com/langgraph) and executed as independent Python services, apply VTS reasoning at ascending scopes while emitting the same evidence-rich JSON schema.

- **Micro observer** that operates at paragraph, table, or figure granularity. For every element it describes the issue, cites the exact words or numbers that triggered the observation, assigns a High/Medium/Low impact score, and, under "What more can we find?", clusters semantically related anomalies across pages to surface latent patterns and root causes.
- **Meso observer** transforms clustered issues into actionable levers. It ingests organizational metadata and maps each problem to a controllable variable, returning quantified targets, implementation steps, resource envelopes, and synergy/ trade-off notes, all explicitly linked back to micro-level evidence.
- **Macro observer** synthesizes multi-quarter documents and meso action sheets into a portfolio strategy. Using J-curve heuristics and resource allocation constraints, it sequences initiatives, schedules capital and headcount ramps, and attaches risk-adjusted NPV projections, each step traceable to the originating snippet in the source filings.

Across all tiers, prompts retain VTS language ("Describe what you notice", "Cite what makes you conclude this", "State additional possibilities") so the model must ground every inference in observable data.

**Result Consolidation with Evidence and Traceability**

Each observation is emitted as a self-contained JSON object that records (i) findings and the exact source location, which contains page number, bounding-box coordinates, and raw excerpt or table cell, (ii) an impact rating, and (iii) any links to organizational parameters or strategic objectives. Downstream modules preserve this metadata verbatim, enabling auditors to walk from a portfolio-level recommendation all the way back to the single sentence or number that triggered it. After the three observers finish processing, their YAML fragments are collated into a rich composite file result.yaml and rendered as result.html for human-readable review.

*Figure 4* presents a concrete example from a recent quarterly financial-report analysis. It shows how the system groups semantically related anomalies under a unique identifier (**PF1** for *Profit/Financial Performance*). For each group it stores a representative issue (e.g. "Negative Net Profit and Declining Profitability") with a plain-language description, a primary page reference, a list of related issues surfaced elsewhere in the report, and an aggregate priority score. This hierarchy lets reviewers zoom fluidly from portfolio themes down to line-item evidence, ensuring that every strategic insights remains fully traceable to what is actually written in the source document.

```yaml
vts_results > Y current_micro_analysis_vts.yaml
grouped_issues:
  Profit/Financial Performance:
    - group_id: PF1
      representative_issue:
        title: Negative Net Profit and Declining Profitability
        description: The company reported a net loss of $668.5K in Quarter 1, with negative
          ROS (-3.5%), low EBITDA margin (3.0%), negative EVA (-$1.41M), and negative
          ROCE (-1.3%) despite a cost of capital of 10.8%. This indicates unprofitable
          operations and inefficient capital use.
        severity: High
        page_reference: '002'
      related_issues:
        - Operating Profit Turnaround and Margin Pressure (page 001)
        - High Interest and Tax Expenses Relative to EBITDA (page 002)
        - Declining Capital Employed and Cost of Capital (page 002)
        - Low Contribution Margin per Account (page 005)
        - High Marketing Investment with Low Market Share in Large Corporate Segment (page
          013)
        - Discount Strategy May Impact Profit Margins in Large Corporate Segment (page
          013)
        - Potential Overwork and Capacity Constraints in TSD Team impacting product development
          (page 007, 017, 028)
        - High Capital Expenditure Requirement for Product Development (page 020)
        - Potential Revenue Risk from Security Concerns (page 024)
        - Loss of Market Share Among Large Corporate Customers (page 025)
        - Pricing Strategy Impact on Small Business Accounts (page 025)
        - Erosion of Large Corporate Market Share (page 026)
        - Poor Performance Against Strategic Objectives (page 027)
        - Decline in Technical Offering Quality (page 028)
        - High Investment Commitment with Unclear ROI (page 031)
      priority_score: 9
    - group_id: PF2
      representative_issue:
        title: Difficulty in Gaining New Accounts and Market Share Loss
        description: The company is losing large corporate accounts and market share
          (pages 003, 025, 026), with stagnant or declining account numbers and market
          share. New account acquisition is difficult, and pricing and marketing strategies
```

**Figure 4. An example of VTS observations**

**Human-in-the-Loop Validation**

Instead of a bespoke web dashboard, we load the original PDFs, the extracted YAML, and the observer outputs directly into an AI-enabled coding IDE (Windsurf AI in our prototype, though Cursor or VS Code are interchangeable). As illustrated in **Figure 5**, the analyst view shows the machine-generated YAML in the main editor pane while Claude-4 sits in a side panel, proposing comparisons, clarifications, or counter-arguments. Analysts accept, amend, or discard each finding; the curated set is then pushed back to the orchestrator for recalibration. This tight review loop marries the reach and consistency of VTS-guided AI with managerial judgment, ensuring that final recommendations remain both evidence-grounded and context-sensitive.

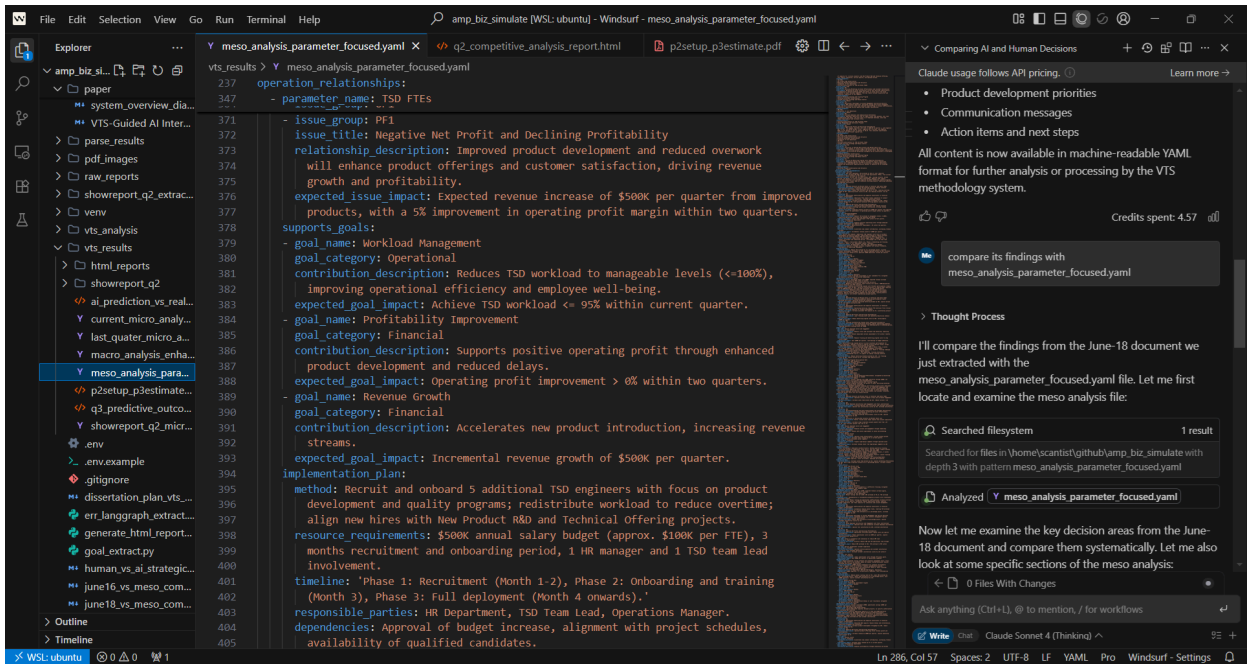

**Figure 5. Human-in-the-Loop Validation and AI Co-Working Interface (Windsurf)**

## Results

To conduct an initial check on the VTS-AI workflow, we ran a small experiment using an 18-page business-simulation report from InsightExperience [16]. The document contains information in various formats, including narrative text, tables, and images. First, we gave the full PDF to ChatGPT-4o with the prompt: "*Read this document page by page, highlight findings related to negative business performance, show evidence, and offer suggestions.*" We then processed the same PDF through our VTS-AI pipeline and compared the two outputs, evaluating how many issues each surfaced, how clearly they cited evidence, and how actionable their recommendations were. This comparison let us gauge whether VTS-AI produces richer, more traceable insights than a single-prompt LLM. Data are available from the authors upon request.

## Comparative Efficiency and Information Density

From an efficiency standpoint, the human analyst typically needs thirty to sixty minutes to absorb the tables, cross-reference figures, and compile notes. ChatGPT finishes its first summary in about a minute, offering a rapid narrative of headline problems. VTS-AI completes its run in roughly the same one-to-two-minute window as ChatGPT because it relies on large-language-model inference as well; however, its internal loop processes the briefing page by page, storing each finding as a discrete data record rather than prose.

Information density tells a different story. Although the human analyst's notes often contain subtle contextual insights, they remain unstructured and difficult to query. ChatGPT's prose is concise but omits page locations, raw metrics, and causal links, so each new analytical angle demands another prompt and another complete pass through the file. VTS-AI achieves the highest density: every YAML line carries the source page, a verbatim quotation or numeric value, a severity score, and a root-cause hint. A later query (e.g.: identifying every indicator of margin compression) is answered instantly from the YAML without reopening the document or re-prompting the model.

Bringing the two dimensions together, the VTS-AI approach occupies the most favorable corner of the trade-off curve. It processes text as fast as sending a prompt directly to ChatGPT. It also preserves, and sometimes even surpasses, the evidence a meticulous human reader would gather. The output is a fully queryable knowledge object. This object can feed dashboards, downstream agents, and audits. Table 2 highlights the core finding: VTS-AI combines near-instant turnaround with a level of traceable detail that neither manual review nor direct prompting can deliver, offering the greatest "information per second" for early-stage document sensing.

**Table 2. Speed and detail captured by review methods: Manual, GPT direct, and VTS-AI**

| Method | Digest Efficiency | Information Density |
|---|---|---|
| Human analyst | 30–60 min | High nuance but unstructured; evidence lives in personal notes and is hard to reuse |
| ChatGPT one-shot prompt | ~1 min | Medium; prose summary lacks page references, raw metrics, and explicit causal mapping |
| VTS-AI pipeline | 1–2 min | Very high; YAML stores page IDs, verbatim figures, severity tags, and cross-domain links |

## Guidance Value

In our business-simulation exercise, we compared VTS-AI's Q3 forecasts with the actual outcomes produced by the InsightExperience simulator. The model's broad signals aligned well with management's expectations:
- it anticipated a further rise in revenue;

- it suggested that EBITDA margins would improve;
- it foresaw higher technology-satisfaction scores; and
- it pointed toward net account growth with reduced churn.

In each case the **trend** was on track, yet the **scale** was off. Revenue came in forty-plus per cent higher than forecast; EBITDA leapt sevenfold beyond the "modest" gain VTS-AI had pencilled in; sales satisfaction scores exploded when the model thought they would stagnate; and new-logo growth ran more than ten times the projected range. The pattern is clear: VTS-AI nails "which way," but without the hidden elasticity curves and conversion rules that power the business simulation, it can only guess at "how far."

These gaps arise from missing context, not faulty logic. The system sees published tables and workload metrics, but it cannot observe the proprietary equations that turn, say, a one-point price cut into a five-point margin lift or a service breakthrough into viral account expansion. Lacking those transfer functions, the model works with proxies, such as historical deltas, utilization ratios, severity scores. So its numbers stay indicative rather than prescriptive.

As a practical takeaway, we should use VTS-AI as an inexpensive first-pass sensor. It is a tool that quickly highlights what deserves attention. Human expertise, detailed financial modeling, and proprietary data must still decide how far to move each dial. The recommended workflow is sequential: run VTS-AI for a rapid, evidence-anchored scan; overlay managerial judgment and cost curves; then fine-tune the targets through a joint human-AI review. This approach blends the reach of automated sensing with the precision of domain knowledge, converting qualitative signals into actionable, quantitative plans. Tables 3 and 4 provide a concise summary of the comparisons.

**Table 3. Financials: Forecast vs. Actual**

| Metric | AI Prediction (range / note) | Q3 Reality | Direction correct? | Size of miss |
|---|---|---|---|---|
| Revenue | $18.5 M ( $18.3 – 18.7 M ) ≈ +3 % growth | $26.24 M | Yes | -$7.7 M (-42 %) |
| EBITDA % | ≈ +1 % (0.5 – 3 %) | 21.1 % | Yes | +18 pts |

**Table 4. Customer Metrics: Predicted vs. Real**

| Metric | AI Prediction | Q3 Reality | Direction correct? | Gap |
|---|---|---|---|---|
| Tech-Offering score | 63 – 67 % | 71 % | Yes | +4 – 8 pts |
| Sales-Satisfaction | 50 – 52 % | 71 % | No | +19 – 21 pts |
| Account-Service score | 76 – 78 % | 71 % | Mixed (lower) | -5 – 7 pts |

Table 5. Account Growth: Expected vs. Achieved

| Metric | AI Prediction | Q3 Reality | Direction correct? | Gap |
|---|---|---|---|---|
| New accounts | 150 – 200 | 128 Large-Corp 4,199 Small-Biz | Yes | > 10 times higher overall |
| Customer churn | 600 – 650 lost | Net and customer base positive growth | No | Full reversal |

# Future Work

Our prototype currently halts at qualitative sensing. The next frontier is to translate those narrative signals into rigorously quantified estimates of value impact, a challenge that sits squarely within the well-established paradigm of cross-sectional analysis [17-20]. Although this paradigm is best known in stock-price prediction, we believe it can enrich VTS-AI as well. To explore that possibility, we lay out the following research questions, which we hope to pursue with finance-domain collaborators.

1. From VTS tags to diagnostic ratios
   Can levers surfaced by VTS, such as "inventory build-up" or "margin squeeze", be mapped onto standard diagnostics like gross-margin, ROIC, or the cash-conversion cycle? We might start with a lightweight mapping table and back-fill the required inputs from publicly available 10-Q filings to test both feasibility and noise.
2. LLM-enhanced analytics via a Model-Context Protocol (MCP)
   If we consider a rigorous financial analysis as a third party service to VTS-AI. How to integrate them correctly and efficiently? Which domain-tuned language models (e.g., FinGPT, BloombergGPT, Llama-3-Fin) and task-specific tools can be cleanly integrated through an MCP layer to deepen analysis beyond tagging? We can wrap finance-focused LLMs and parsers (FinGPT for Q & A filings Q&A, PyMeritrage for options data, yFinance feeds, etc.) as callable MCP tools. We can also test agent workflows that chain VTS tags, MCP-exposed models/tools and structured JSON outputs (ratios, sentiment scores, forward-guidance vectors).

A dedicated Risk & Safety Layer must operate alongside the core VTS-AI engine. On the financial-assurance front, the layer will stress-test every valuation formula and ratio for input sensitivity, automatically flag missing or restated 10-Q data before it reaches the pipeline, and attach plain-language rationales with complete parameter logs so regulators and users can audit any forecast. It will enforce strict token and API budgets to keep cloud costs predictable, mandate human sign-off for high-stakes recommendations, and guard against single-provider outages by routing language models and data feeds through interchangeable interfaces.

Cyber-resilience is equally critical. The design will encrypt all stored data, limit log retention, and purge sensitive prompts on a fixed schedule. Incoming text will be scanned for malicious patterns to block prompt-injection attacks, while every model weight and library will be verified against a signed registry. All filings will be cross-checked against multiple sources to detect forgeries, third-party API keys will be tightly scoped and rotated, and the platform will undergo denial-of-service drills and red-team exercises to prove it fails safely.

## Conclusion

Embedding Visual Thinking Strategies (VTS) prompts into an AI-mediated interview workflow enriches qualitative insights while streamlining analysis. The study demonstrates that pairing disciplined human inquiry with large-language-model support surfaces a broader range of themes in less time than unguided chat alone, without compromising reliability. Future work discusses how to extend the framework to multilingual settings, larger participant pools, and automated linkage of emergent themes to organizational knowledge graphs, moving toward a fully integrated, AI-assisted insight pipeline.

## Acknowledgments

All authors participated in the 2025 MIT Advanced Management Program [21] and contributed equally to this paper. The study originated in that program's blend of leadership practice, disciplined methodology, and AI-driven analysis. We gratefully acknowledge the MIT Sloan faculty and staff for their meticulous planning and tireless support; Insight Experience for the business-simulation testbed; Dabney Hailey [22] for her guidance on Visual Thinking Strategies; and our MIT AMP classmates for their generous feedback and encouragement.

## References

[1] Vaswani, A., et al. "Attention Is All You Need." Advances in Neural Information Processing Systems, 2017.
[2] "LLM in Finance: Revolutionizing Financial Operations." Debut Infotech Blog, 2024.
[3] Wu, S., et al. "BloombergGPT: A Large Language Model for Finance." arXiv preprint arXiv:2303.17564, 2023.
[4] Peng, Z., et al. "DocLLM: A Unified Framework for Document Understanding with Large Language Models." arXiv preprint arXiv:2401.07720, 2024.
[5] Peng, Z., et al. "Large Language Models for Financial and Investment Management." MIT Media Lab, 2024.
[6] "Building AI Capabilities Into Portfolio Companies at Apollo." MIT Sloan Management Review, 2025.
[7] "AI Agents in Finance: How Agentic AI is Powering the Next Generation of FP&A." FPA Trends, 2024.
[8] Gou, S., et al. "Multi-Agent System for Financial Numerical Reasoning." ACL Anthology, 2024.
[9] "4 new studies about agentic AI from the MIT Initiative on the Digital Economy." MIT Sloan School of Management, 2025.
[10] "AI Agents in Finance: A Game-Changer for Financial Services." DigiQT Blog, 2024.

[11] Kim, A. "Financial Statement Analysis with Large Language Models." Bayes Business School, 2024.

[12] "Ensuring Explainability and Auditability in Generative AI Copilots for Fincrime Investigations." Lucinity Blog, 2023.

[13] Lewis, P., et al. "Retrieval-Augmented Generation for Knowledge-Intensive NLP Tasks." Advances in Neural Information Processing Systems, 2020.

[14] "The Power of AI Agents for Finance Due Diligence." Pathway Blog, 2024.

[15] "Generative AI in Finance: Use Cases and Future Trends." V7 Labs Blog, 2024.

[16] Insight Experience Business Simulation

https://www.insight-experience.com/case-studies/developing-effective-global-leaders-a-case-study

[17] Asgharian, H. and Hansson, B. "Cross Sectional Analysis of the Swedish Stock Market." Lund University Working Paper, 2002.

[18] Greenwood, R. M. "A Cross-Sectional Analysis of the Excess Comovement of Stock Returns." SSRN Working Paper, 2005.

[19] Wang, L. "Deep Cross-Sectional Stock-Return Prediction with Neural Networks." *Journal of Financial Data Science*, 2024.

[20] Chen, Y., Li, X. and Zhang, J. "Testing Factor Models under Cross-Stock Dependence." *Review of Asset Pricing Studies*, 2025.

[21] MIT Sloan School of Management, "Advanced Management Program", https://executive.mit.edu/course/advanced-management-program

[22] D. Hailey, "LinkedIn profile," *LinkedIn*. https://www.linkedin.com/in/dabneyhailey/